\documentclass[%
 reprint,
 amsmath,amssymb,
 aps,
]{revtex4-2}

\usepackage{graphicx}
\usepackage{dcolumn}
\usepackage{bm}
\usepackage{xcolor}

\begin{document}

\title{
Constructing efficient strategies for the process optimization by restart}

\author{Ilia Nikitin and Sergey Belan}
\email{sergb27@yandex.ru}
\affiliation{Landau Institute for Theoretical Physics, Russian Academy of Sciences, 1-A Akademika Semenova av., 142432 Chernogolovka, Russia}
\affiliation{National Research University Higher School of Economics, Faculty of Physics, Myasnitskaya 20, 101000 Moscow, Russia}

\begin{abstract} 
Optimization of the mean completion time of random processes by restart is a subject of active theoretical research in statistical physics and has long found practical application in computer science. 
Meanwhile, one of the key issues remains largely unsolved: how to construct a restart strategy for a process whose detailed statistics are unknown to ensure that the expected completion time will reduce? 
Addressing this query here we propose several constructive criteria for the effectiveness of various protocols of non-instantaneous restart in the mean completion time problem and in the success probability problem. Being expressed in terms of a small number of easily estimated statistical characteristics of the original process (MAD, median completion time, low-order statistical moments of completion time), these criteria allow informed restart decision based on partial information.
\end{abstract}
\maketitle

\vskip \baselineskip

\section{Introduction}
Restart as a method of accelerating randomized tasks was first proposed in the early 90s in computer science. 
Namely, the authors of Refs. \cite{Alt_1991, Luby_1993} showed that applying a restart to a probabilistic algorithm whose completion time represents a highly fluctuating random variable leads to smaller tail probabilities and to smaller expected completion time. 
Beyond computer science, optimization via restart is the subject of active research by the statistical physics community. 
The starting point for these studies was the work of Evans and Majumdar \cite{EM_2011}, who found that stochastic (Poisson) restart reduces the mean first-passage time of Brownian motion. 
Also, this area of research received an additional impetus for development due to the works \cite{Reuveni_2014, Rotbart_2015, Reuveni_PRL_2016} devoted to the kinetics of enzymatic reactions, where the restart corresponds to dissociation of the intermediate enzyme-substrate complex.
Theoretical analysis presented in these studies demonstrated that an increase in the dissociation rate constant could potentially speed up enzymatic turnover.
 
Thus, essentially the same mathematical model arises in different research fields. Regardless of the specific context, one of the critical tasks is the following: how to find a restart strategy that is guaranteed to reduce the expected completion time of the stochastic process of interest?
The general renewal formalism developed by Pal and Reuveni \cite{Reuveni_PRL_2017} allows to predict whether the implementation of a particular restart protocol will be effective for a given stochastic process. 
Moreover, as proved in \cite{Luby_1993,Lorenz_2021,Reuveni_PRL_2017} (see also \cite{pal2016diffusion}), a strictly regular (periodic) strategy, implying that
the random process is restarted every $\tau_\ast$ units of time, is universally optimal.
The value of the optimal period $\tau_\ast$ is problem-specific and can be determined once we know the probability density of the random completion time of the original process.

In practice, however, the statistics of the optimized process can be poorly specified or even completely unknown \cite{Luby_1993,Gagliolo_2007,Reuveni_2014,Lorenz_2018,Lorenz_2021,Wu_2007,Streeter_2007}. 
In the recent paper \cite{Eliazar_JPA_2021}, which addresses a scenario of partially known statistics,
Eliazar and Reuveni formulated the \textit{constructive criteria} for the effectiveness of restarts, expressed through the simple statistical characteristics of a randomized task. 
Unlike previously known \textit{existence results} \cite{Reuveni_PRL_2016, starkov2023universal}, constructive criteria serve not just as indicators of the restart effectiveness, but also offer a strategy that is guaranteed to reduce the average completion time.

The results presented in \cite{Eliazar_JPA_2021}, however, are obtained under the assumption of instantaneous restart events, while in real-life settings, a restart is accompanied by some time delays.
Say, in the context of single-molecule enzyme kinetics, some time is required for the enzyme that unbinds from the substrate to find a new one in the surrounding solution \cite{Reuveni_2014,Reuveni_PRL_2016,Robin_2018a}. Similarly, the restart of a computer program typically involves a time overhead. Also, models with non-instantaneous restarts provide more realistic pictures of colloidal particle diffusion with resetting \cite{Friedman_2020a}. To the best of our knowledge, the existing literature lacks constructive criteria of restart efficiency for models with non-momentary restarts.
In addition, the criteria proposed in \cite{Eliazar_JPA_2021} refer only to the case of a periodic restart. 
The constructive criteria, if any, for stochastic restart strategies remain unknown.

Finally, let us note that the potential of the restart is not limited to optimization of the mean completion time. 
In particular, the implementation of restart  can also improve the probability of getting a desired outcome when a process has several alternative completion scenarios \cite{Belan_PRL_2018}.
Constructive criteria for the problem of optimizing the splitting probabilities have not yet been formulated as far as we know.


In this paper, we fill the aforementioned gaps by constructing a set of constructive efficiency criteria for (i) periodic non-instantaneous restarts and (ii) stochastic non-instantaneous restarts. In addition to the mean completion time problem, we analyze the success probability optimization.
The resulting criteria specify the range of effective values of the control parameter of a periodic or stochastic restart strategy through such characteristics as low-order statistical moments, median value, and mean absolute deviation of the random completion time in the absence of the restart.

\section{Optimization of the mean completion time}
Consider a stochastic process whose random duration time $T$ has a probability density $P(T)$.
The restart protocol ${\cal R}$ is characterized by a (possibly infinite) sequence of inter-restart time intervals $\tau_1,\tau_2,\dots$.
If the process is completed prior to the first restart event, the story ends there. 
Otherwise, the current attempt is aborted, and the process starts from scratch. 
Similarly, the next attempt may either complete prior to the second restart or not, with the same rules. 
This procedure repeats until the process finally reaches completion. 
We will also assume that the initialization of the process and each restart are accompanied by some random time delay $t$  which is independent of $T$. 

In the absence of the restart, the random waiting time for the process completion is given by the sum  $T+t$.
The restart protocol is considered effective if $\langle T_{{\cal R}}\rangle<\langle T\rangle+\langle t\rangle$, where 
$T_{{\cal R}}$ is a random waiting time for  the process completion in the presence of the protocol ${\cal R}$, and angular brackets denote averaging over the statistics of the original process and possibly over the statistics of the inter-restart intervals (in the case of a stochastic protocol).
It is convenient to define the dimensionless effectiveness of the restart as 
\begin{equation}
\label{efficiency_def}
\eta_{{\cal R}}=1-\frac{\langle T_{{\cal R}}\rangle}{\langle T\rangle + \langle t\rangle}.
\end{equation}
Clearly, the effective protocols obey $0<\eta_{{\cal R}}\le 1$.





In the simplest case of a strictly regular schedule, implying that the restart events are equally spaced in time, i.e., $\tau_k=\tau$ for all $k=1,2,...$, the expected completion time can be expressed as (see Appendix \ref{appendix1})
\begin{equation}\label{completion_time}
    \langle T_{\tau}\rangle=\frac{\langle T\rangle+\tau-\langle |T-\tau|\rangle+2\langle t\rangle}{2Pr[T\le\tau]},
\end{equation}
where $\langle |T-\tau|\rangle$ is the mean absolute deviation (MAD) of the random variable $T$ from the value of $\tau$. 
Although, the expression for the mean completion time under regular restart has long been known \cite{pal2016diffusion,Reuveni_PRL_2017}, its particular representation (\ref{completion_time}) in terms of MAD was first obtained relatively recently in Ref. \cite{Eliazar_JPA_2021} for the particular case of zero time penalty $t=0$.

Another important scenario is the Poisson restart, where inter-restart intervals are mutually independent and identically distributed with exponential probability density $\rho(\tau)=r e^{-r\tau}$.
Here $r$ represents the restart rate. The average completion time in the presence of a Poisson restart is given by (see \cite{Reuveni_PRL_2016})
\begin{equation}\label{eq:T_r_mean_main}
    \langle T_r\rangle=\frac{1-\tilde{P}(r)+r\langle t\rangle}{r\tilde{P}(r)},
\end{equation}
where $\tilde{P}(r)=\int_0^\infty dT P(T)e^{-rT}$ denotes the Laplace transform of the probability density function $P(T)$.

In addition, we will consider the case of a gamma protocol for which the random intervals between restarts are independently sampled from the Gamma distribution $\rho(\tau)=\frac{\beta^{k}}{\Gamma(k)}\tau^{k-1} e^{-\beta \tau}$  with rate parameter $\beta$ and shape parameter $k$. 
The expected completion time of a process under gamma restart with $k = 2$ is given by (see Appendix \ref{appendix2})
\begin{equation}\label{mean_T_Gamma}
    \langle T_{\beta}\rangle=\frac{\beta\langle t\rangle+2-2\tilde P(\beta)+\beta\partial_\beta\tilde P(\beta)}{\beta\tilde P(\beta)-\beta^2\partial_\beta \tilde P(\beta)}.
\end{equation}

Let us stress that $r$ entering Eq. (\ref{eq:T_r_mean_main}) can be interpreted both as the parameter of exponential distribution and as the constant rate of restart events.  
On the contrary, since gamma-distributed variables are not memory-less, $\beta$ in Eq. (\ref{mean_T_Gamma}) represents just a parameter of gamma distribution, while the rate of restart events in this case is time-dependent $\bar \beta(t) = \frac{\beta^{\kappa}\tau^{\kappa-1}e^{-\beta\tau}}{\Gamma(\kappa,\beta \tau)}$, see, e.g., \cite{Reuveni_PRL_2017}. 
At the same time, $\beta$ determines the mean period between restart events, which is equal to $2\beta^{-1}$ when $k=2$.

Once full statistics of the original process encoded into the probability density $P(T)$ are known, one can use Eqs. (\ref{completion_time})-(\ref{mean_T_Gamma}) to determine the values (if any) of the control parameter ($\tau$, $r$ or $\beta$) for which the corresponding strategy is advantageous.
Moreover, the most efficient strategy is always periodic, and the optimal period can be found by minimizing the right side of Eq. (\ref{completion_time})  with respect to $\tau$.
Our goal is to formulate constructive criteria of restart efficiency, allowing to choose a guaranteed beneficial restart protocol without knowledge of the full process statistics. Several such criteria were proposed in \cite{Eliazar_JPA_2021} for the particular case of the periodic restart with zero penalty $t=0$.

Let us outline the idea underlying our derivation of the desired criteria. 
First of all, using various probabilistic inequalities, we obtain an upper bound for the mean completion time $\langle T_{{\cal R}}\rangle\le{\cal T}$ of the process under restart protocol ${\cal R}$, where the time scale ${\cal T}$ is expressed through  the control parameter, which determines the mean restart period ($\tau$, $r$ or $\beta$ depending on the protocol), and some simple statistical characteristics of the original process (e.g., statistical moments, median value, MADs, etc.). 
Then the solution of the inequality ${\cal T}\le\langle T \rangle +\langle t\rangle$ with respect to the mean period defines a set (possibly empty) of efficient values for period. Further, by minimizing ${\cal T}$ on this set, one can find the mean restart period that provides the maximum guaranteed efficiency among the specified effective values of the control parameter.
Next, we demonstrate concrete implementations of this simple idea.

\subsection{Periodic restart}

For any random variable $T$ one has $\langle |T-\tau|\rangle\ge \langle |T-m|\rangle$, where $m$ is the median value of $T$.
Moreover, equality is achieved for $\tau=m$.
So, using Eq. (\ref{completion_time}), we get the following estimate
\begin{equation}
\label{eq5}
\langle T_{\tau}\rangle\le  \frac{\langle T\rangle+\tau-\langle |T-m|\rangle+2\langle t\rangle}{2Pr[T\le\tau]}.
\end{equation}
It follows from Eq. (\ref{eq5}) that if the condition 
\begin{eqnarray}
\label{eq6}
    m + \langle t\rangle < \langle |T-m|\rangle,
\end{eqnarray}
is satisfied, then 
for any $\tau$ belonging to the interval
\begin{eqnarray}
\label{eq6_0}
m\le\tau< \langle |T-m|\rangle - \langle t\rangle,
\end{eqnarray}
the  inequality $\langle T_{\tau}\rangle\le  \langle T\rangle+\tau-\langle |T-m|\rangle + 2\langle t\rangle<\langle T\rangle + \langle t\rangle$ holds.
Thus, Eq.  (\ref{eq6}) represents the sufficient condition for the existence of the effective periodic restart strategy, and Eq. (\ref{eq6_0}) gives a recommendation regarding the particular values of the restart period.  
It is easy to see that the choice $\tau_0=m$ is optimal in the sense that regular restart with period $\tau_0$ provides the highest guaranteed efficiency among the values specified by Eq. (\ref{eq6_0}), which is given by
\begin{equation}
\eta_1 = \frac{\langle |T-m|\rangle - m - \langle t\rangle}{\langle T \rangle + 
 \langle t \rangle}.
\end{equation}
The described criterion generalizes the result presented \cite{Eliazar_JPA_2021} to the case of a non-zero duration of restart events.

Further, as shown in \cite{starkov2023universal} for the particular case $t=0$, the inequality $2m<\langle T\rangle$ represents a sufficient condition for the existence of an effective periodic protocol.
Here we turn this existence result into a constructive criterion that holds in the presence of a restart penalty.
By virtue of Jensen's inequality, one has $\langle |T-\tau|\rangle\ge\langle T\rangle-\tau$, and, therefore, Eq. (\ref{completion_time}) allows to write the following estimate 
\begin{equation}
\label{eq7}
\langle T_{\tau}\rangle\le \frac{\tau + \langle t \rangle}{Pr[T\le \tau]}.
\end{equation}
From Eq. (\ref{eq7}) we see that if the condition
\begin{eqnarray}
\label{eq10}
    m< \frac12\langle T\rangle - \frac12\langle t\rangle,
\end{eqnarray}
is met, then regular restart with the period belonging to the interval  
\begin{eqnarray}
    m\le\tau< \frac12\langle T\rangle-\frac12\langle t\rangle,
\end{eqnarray}
reduces the average completion time since in this case one gets $\langle T_{\tau}\rangle\le 2\tau+2\langle t\rangle<\langle T\rangle +\langle t\rangle$.
The guaranteed efficiency is maximal at $\tau_0=m$ and is estimated as
\begin{equation}
\eta_2 \ge 1 - \frac{2 (m + \langle t \rangle)}{ \langle T \rangle + \langle t \rangle}.
\end{equation}

\subsection{Poisson restart}
Periodic strategy is important due to its optimal property \cite{Luby_1993,Lorenz_2021,Reuveni_PRL_2017}, which has already been discussed above. Namely, if you found a value $\tau_{\ast}\ge0$ (probably $\tau_{\ast}=+\infty$) such that $\langle T_{\tau_{\ast}}\rangle\le \langle T_{\tau}\rangle$ for any $\tau\ge 0$, then $\langle T_{\tau_{\ast}}\rangle\le \langle T_{\cal R}\rangle$ for all ${\cal R}$. 
However, as previous studies have shown \cite{Reuveni_PRL_2016}, a periodic protocol with a non-optimal period $\tau\ne\tau_\ast$ may be inferior to other restart strategies. 
Since the optimal period $\tau_\ast$ of a regular restart does not have to be equal to the median completion time of the initial process $m$, the periodic strategies constructed above are generally not optimal. 
In addition, the conditions of their applicability (see Eqs. (\ref{eq6}) and (\ref{eq10})) are not necessary for the existence of an effective protocol.
In other words, violation of the conditions $m+\langle t\rangle< \langle |T-m|\rangle$ and $m< \frac12\langle T\rangle-\frac12\langle t\rangle$ for a given stochastic process does not mean that an effective restart protocol cannot be constructed.

Given above arguments, it is interesting to develop efficient policies of non-periodic restarts. 
Particularly attractive in this respect is the Poisson strategy, which has been widely studied before, see e.g. \cite{EM_2011, Reuveni_2014, Rotbart_2015, Reuveni_PRL_2016,Pal_JPA_2022,evans2020stochastic}. 
A simple sufficient condition for the efficiency of the Poisson restart \cite{Pal_JPA_2022}
reads
\begin{eqnarray}
\label{T_r_pois_sc}
\langle T^2\rangle>2\langle T\rangle(\langle T\rangle+\langle t\rangle).
\end{eqnarray} 
Note, however, that the latter inequality represents the existence result: it serves as an indicator of the existence of an effective Poisson strategy without presenting it. Moreover, the logic underlying the derivation of the sufficient condition determined by Eq. (\ref{T_r_pois_sc}) suggests that knowing the first two statistical moments of the random completion time is not enough to choose an effective restart rate.
Namely,  the pair $\langle T\rangle$ and $\langle T^2\rangle$ determines only the slope of  $\langle T_r\rangle$ in its dependence on $r$ at the point $r=0$, saying nothing about its behavior for non-zero values of $r$.

A constructive condition for the effectiveness of the Poisson restart can be formulated if we add information about the third-order moment. 
Based on the knowledge of $\langle T\rangle$, $\langle T^2\rangle$, and $\langle T^3\rangle$, one can estimate the Laplace transform of  $P(T)$ as  (see \cite{Zubkov_1999} and Appendix \ref{appendix3})
\begin{equation}\label{eq:P_r_mod_main}
    \tilde{P}(r) \ge 1 - r \langle T \rangle + \frac{r^2 \langle T^2 \rangle^2}{r\langle T^3 \rangle + 2 \langle T^2 \rangle}.
\end{equation}
Next, assuming that the expression on the right side of \eqref{eq:P_r_mod_main} is positive, from \eqref{eq:T_r_mean_main} and \eqref{eq:P_r_mod_main} we get the following estimate


\begin{equation}
\label{estimate_T_r_main}
    \langle T_r\rangle \le \frac{(\langle t\rangle+\langle T\rangle)(r \langle T^3 \rangle + 2 \langle T^2 \rangle) - r {\langle T^2 \rangle}^2}{(1 - r\langle T\rangle)(r \langle T^3 \rangle + 2 \langle T^2 \rangle) + r^2 {\langle T^2 \rangle}^2}.
\end{equation}

Therefore, if $r$ satisfies the set of inequalities
\begin{equation}
\left\{\begin{array}{ll}
\frac{(\langle t\rangle+\langle T\rangle)(r \langle T^3 \rangle + 2 \langle T^2 \rangle) - r {\langle T^2 \rangle}^2}{(1 - r\langle T\rangle)(r \langle T^3 \rangle + 2 \langle T^2 \rangle) + r^2 {\langle T^2 \rangle}^2}<\langle t\rangle+\langle T\rangle, \\
\\
\color{black}
\label{eq:r_cond_mod}
 1 - r\langle T\rangle + \frac{r^2 {\langle T^2 \rangle}^2}{r \langle T^3 \rangle + 2 \langle T^2 \rangle}>0,\\
 \\
 r>0,
\end{array} \right.
\end{equation}
then the Poisson restart at rate $r$ reduces the mean completion time. 
Solving this system under the assumption that a sufficient condition defined by \eqref{T_r_pois_sc} is met, we find an interval of effective rates  
\begin{eqnarray}
\label{effective_r_m}
0 < r < \frac{\langle T^2 \rangle (\langle T^2 \rangle -2 \langle T \rangle  (\langle T\rangle+\langle t \rangle))}{(\langle T\rangle+ \langle t \rangle)(\langle T \rangle \langle T^3 \rangle-\langle T^2 \rangle^2)}.
\end{eqnarray}

Further, within the interval given by \eqref{effective_r_m}, the expression on the right side of the inequality \eqref{estimate_T_r_main} attains its minimum value at the point $r_3$ given by Eq. (\ref{eq:r^m_opt_main}) in Appendix \ref{appendix5}. 
As can be found from (\ref{effective_r_m}), (\ref{eq:P_r_mod_main}) and (\ref{estimate_T_r_main}), the resulting efficiency is given by (\ref{eq:eta_3}), see Appendix \ref{appendix5}.

In the limit of negligible restart duration $t=0$, we obtain fairly compact expressions for the optimal point
\begin{equation}\label{eq:r^m_opt_main_limit}
    r_{0} = \frac{\sqrt{2} \langle T^2 \rangle ^{3/2}-2\langle T \rangle \langle T^2 \rangle }{\langle T \rangle \langle T^3 \rangle - \langle T^2 \rangle^2},
\end{equation}
while the corresponding estimate for dimensionless effectiveness reads
\begin{eqnarray}
\label{eta_3}
    \eta_3 \ge \frac{2 \langle T^2 \rangle \langle T \rangle^2+\langle T^2 \rangle^2-2 \sqrt{2}\langle T \rangle \sqrt{\langle T^2 \rangle^3}}{\langle T \rangle \left(2 \langle T \rangle \langle T^2 \rangle+\langle T^3 \rangle-2 \sqrt{2} \sqrt{\langle T^2 \rangle^3}\right)}.
\end{eqnarray}


%
%

Note that the proposed method of constructing an effective Poisson strategy is one of many possible. Alternatively, the Laplace transform of the probability density $P(T)$ can be estimated as \cite{Zubkov_1999} 
\begin{eqnarray}
\label{eq:P_r_classic_main}
  \tilde P(r)\ge  \sum_{k=0}^{2l-1} (-1)^k \frac{r^k \langle T^k \rangle}{k!}, 
\end{eqnarray}
where $l=1,2,...$.
Let us evaluate the term $\tilde P(r)$ in the numerator of the right-hand side of the formula \eqref{eq:T_r_mean_main} using \eqref{eq:P_r_classic_main} with $l=2$, i.e. $\tilde{P}(r)\ge 1 - r\langle T\rangle+\frac{r^2}{2}\langle T^2\rangle-\frac{r^3}{6}\langle T^3\rangle$. Next, to estimate the Laplace transform $\tilde P(r)$ entering the denominator in the same expression, we use Eq. \eqref{eq:P_r_classic_main} with $l=1$: $\tilde{P}(r)\ge 1 - r\langle T\rangle$. 
Then, exploiting these estimates and assuming that $r<1/\langle T\rangle$ we get from Eq. (\ref{eq:T_r_mean_main})
\begin{equation}
\label{estimate_T_r_classic}
    \langle T_r\rangle \le \frac{6 \langle t \rangle +6\langle T\rangle- 3r\langle T^2 \rangle + r^2\langle T^3 \rangle}{6 - 6 r\langle T\rangle},
\end{equation}
and, thus, the interval of effective rates is determined by the following set of inequalities
\begin{equation}
\left\{\begin{array}{ll}
\frac{6\langle t\rangle +6\langle T\rangle - 3r\langle T^2 \rangle + r^2\langle T^3 \rangle}{6 - 6 r\langle T\rangle}<\langle t\rangle+\langle T\rangle, \\
\\
\label{set00}
r<\frac{1}{\langle T\rangle},\\
 \\
 r>0.
\end{array} \right.
\end{equation}
Assuming that the condition (\ref{T_r_pois_sc}) is fulfilled, we readily find a solution
\begin{eqnarray}
\label{T_r_range_standart}
    0<r< \frac{3(\langle T^2 \rangle - 2 \langle T \rangle (\langle T\rangle+\langle t \rangle))}{\langle T^3 \rangle},
\end{eqnarray}
which obviously differs from that given by Eq. (\ref{effective_r_m}) since its derivation is based on a different estimate of $\tilde P(r)$.
Rate $r_4$ determined by Eq. (\ref{eq:r^c_opt_main}) in Appendix \ref{appendix5}
minimizes the right side of the inequality (\ref{estimate_T_r_classic}) on the interval (\ref{T_r_range_standart}), thus providing the highest guaranteed efficiency whose estimate from below is given by Eq. (\ref{eq:eta_4}).

Finally, using \eqref{eq:P_r_mod_main} to estimate the Laplace transform in the numerator of the right side of the formula (\ref{eq:T_r_mean_main}) and the weaker inequality $\tilde{P}(r) \ge 1 - r\langle T\rangle$, which follows from (\ref{eq:P_r_mod_main}), to estimate the denominator, we get a third way to build an effective Poisson protocol. Assuming $r<1/\langle T\rangle$, for the average completion time with this approach, we have
\begin{equation}
\label{estimate_T_r_mod_simple}
    \langle T_r\rangle < \frac{2 (\langle t\rangle+\langle T\rangle)  \langle T^2 \rangle + r \left[(\langle t\rangle+\langle T\rangle)  \langle T^3 \rangle- \langle T^2 \rangle^2 \right]}{(1 -r \langle T \rangle) \left(r \langle T^3 \rangle+2 \langle T^2 \rangle\right)}.
\end{equation}
A set of guaranteed effective rates is determined by the solution of the system of inequalities
\begin{equation}
\left\{\begin{array}{ll}
\frac{2(\langle t\rangle+\langle T\rangle)  \langle T^2 \rangle + r \left[(\langle t\rangle+\langle T\rangle) \langle T^3 \rangle- \langle T^2 \rangle^2 \right]}{(1 -r \langle T \rangle) \left(r \langle T^3 \rangle+2 \langle T^2 \rangle\right)}<\langle t\rangle+\langle T\rangle, \\
\\
\label{eq:r_cond_mod_simple}
 r<\frac{1}{\langle T\rangle},\\
 \\
 r>0,
\end{array} \right.
\end{equation}
solving which we get
\begin{eqnarray}
    \label{interval_mod_simple}
    0<r<\frac{\langle T^2 \rangle \left[\langle T^2 \rangle-2\langle T \rangle (\langle t\rangle+\langle T\rangle)\right]}{(\langle t\rangle+\langle T\rangle)  \langle T \rangle \langle T^3 \rangle}.
\end{eqnarray}

The optimal rate and the corresponding efficiency are determined by
Eqs. (\ref{eq:r^ms_opt_main}) and (\ref{eq:eta_5}) in Appendix \ref{appendix5}.



More generally, using the inequality (\ref{eq:P_r_classic_main}) for $l>2$, as well as combining it with (\ref{eq:P_r_mod_main}), it is possible, in principle, to obtain an unlimited number of other guaranteed effective Poisson strategies. Here we have given only the simplest protocols that require knowledge of a minimum number of statistical moments for their application.


\subsection{Gamma strategy}
Let us now analyze the stochastic gamma protocol. From (\ref{mean_T_Gamma}), we find that if the condition 
\begin{eqnarray}
\label{gamma_sc}
    \langle T^{3}\rangle\ge 3 (\langle T\rangle+\langle t\rangle)\langle T^2\rangle,
\end{eqnarray} 
is met then $\partial_\beta \langle T_\beta\rangle<0$ at $\beta=0$.
So, the inequality (\ref{gamma_sc}) represents a sufficient condition for the existence of an effective gamma strategy. However, the knowledge of the first three moments, $\langle T\rangle$, $\langle T^{2}\rangle$, and $\langle T^{3}\rangle$, does not allow us to choose a value of the parameter $\beta$ that reduces the mean completion time for sure. A desired constructive criterion for the effectiveness of the gamma strategy can be formulated by adding information about the fourth-order statistical moment $\langle T^4\rangle$.

To work out an upper estimate for the average completion time $\langle T_\beta\rangle$ given by (\ref{mean_T_Gamma}), it is useful to note that the derivative of the Laplace transform $\partial_\beta\tilde P(\beta)$ can be represented as 
\begin{eqnarray}
\label{der_Laplace}
    &&\partial_\beta\tilde P(\beta)=\partial_\beta\int_0^\infty dT P(T)e^{-\beta T}=\\
    &&=-\int_0^\infty dT P(T)Te^{-\beta T}=-\langle T\rangle \tilde Q(\beta),
\end{eqnarray}
where $Q(T)=\frac{T}{\langle T\rangle}P(T)$. Being non-negative and normalized by unity, $Q(T)$ can be treated as a probability density of some random variable, and, therefore, its Laplace transform $\tilde Q(r)$ can be evaluated from below using the inequality (\ref{eq:P_r_classic_main})  
\begin{eqnarray}
\label{estimate_for_Q}
  \tilde Q(\beta)\ge  \sum_{k=0}^{2l-1} (-1)^k \frac{\beta^k \langle T^k \rangle_Q}{k!}, 
\end{eqnarray}
where $\langle T^n\rangle_Q\equiv\int_0^\infty T^n Q(T)dT=\frac{\langle T^{n+1}\rangle}{\langle T\rangle}$. From (\ref{der_Laplace}) and (\ref{estimate_for_Q})  we then find
\begin{equation}\label{eq:der_P_r}
   \partial_\beta \tilde{P}(\beta) \le \sum_{k=0}^{2l-1}\frac{(-1)^{k+1}}{k!}\beta^k\langle T^{k+1}\rangle.
\end{equation}


Now let us estimate the term $\tilde{P}(\beta)$ in the numerator of Eq. \eqref{mean_T_Gamma} via the inequality (\ref{eq:P_r_classic_main}) at $l = 2$.
Also, let us use (\ref{eq:P_r_classic_main}) at $l = 2$ to estimate the term $\tilde{P}(\beta)$ in the denominator.
 The term $\partial_\beta\tilde P(\beta) = -\langle T\rangle \tilde Q(\beta)$ in the denominator of Eq. \eqref{mean_T_Gamma} can be evaluated using (\ref{estimate_for_Q}) with $l = 1$.
Finally, in the numerator of Eq. (\ref{mean_T_Gamma}), we use (\ref{estimate_for_Q}) with $l = 2$ for $\partial_\beta\tilde P(\beta)$. 
Combining all these estimates, we obtain

\color{black}
\begin{eqnarray}
\label{T_beta_bound}
    \langle T_\beta\rangle\le \frac{6\langle t\rangle + 6\langle T\rangle-\beta^2\langle T^3\rangle+{\beta^3}\langle T^4\rangle}{6-3{\beta^2}\langle T^2\rangle-{\beta^3}\langle T^3\rangle},
\end{eqnarray}
which implies that $1-\frac{\beta^2}{2}\langle T^2\rangle-\frac{\beta^3}{6}\langle T^3\rangle>0$.
Then the effective parameters $\beta$ can be found from the requirement that
the right-hand side of Eq. (\ref{T_beta_bound}) is less than the mean completion time in the absence of restart $\langle T\rangle+\langle t\rangle$.
This yields the following interval
\begin{eqnarray}
\label{yet_another_range}
    0<\beta<\frac{\langle T^3\rangle-3(\langle T\rangle + \langle t\rangle) \langle T^2\rangle}{(\langle T\rangle +\langle t\rangle)\langle T^3\rangle+\langle T^4\rangle}.
\end{eqnarray}
As expected, this interval (\ref{yet_another_range}) is not empty if the existence conditions (\ref{gamma_sc}) are met.
The best rate parameter providing maximum guaranteed efficiency among the values specified by Eq. (\ref{yet_another_range}) can be found by solving an algebraic equation of the third degree.

Following the same logic, one can derive the effective gamma strategy with an arbitrary natural shape parameter $k$ of the probability density of inter-restart intervals $\rho(\tau)=\frac{\beta^k}{\Gamma(k)}\tau^{k-1}e^{-\beta \tau}$. Note, however, that the larger $k$ is, the greater the number of statistical moments is required to determine the range of effective values of the rate parameter $\beta$.

\begin{widetext}
\begin{center}
    \begin{figure}
        \includegraphics[width=0.7\textwidth]{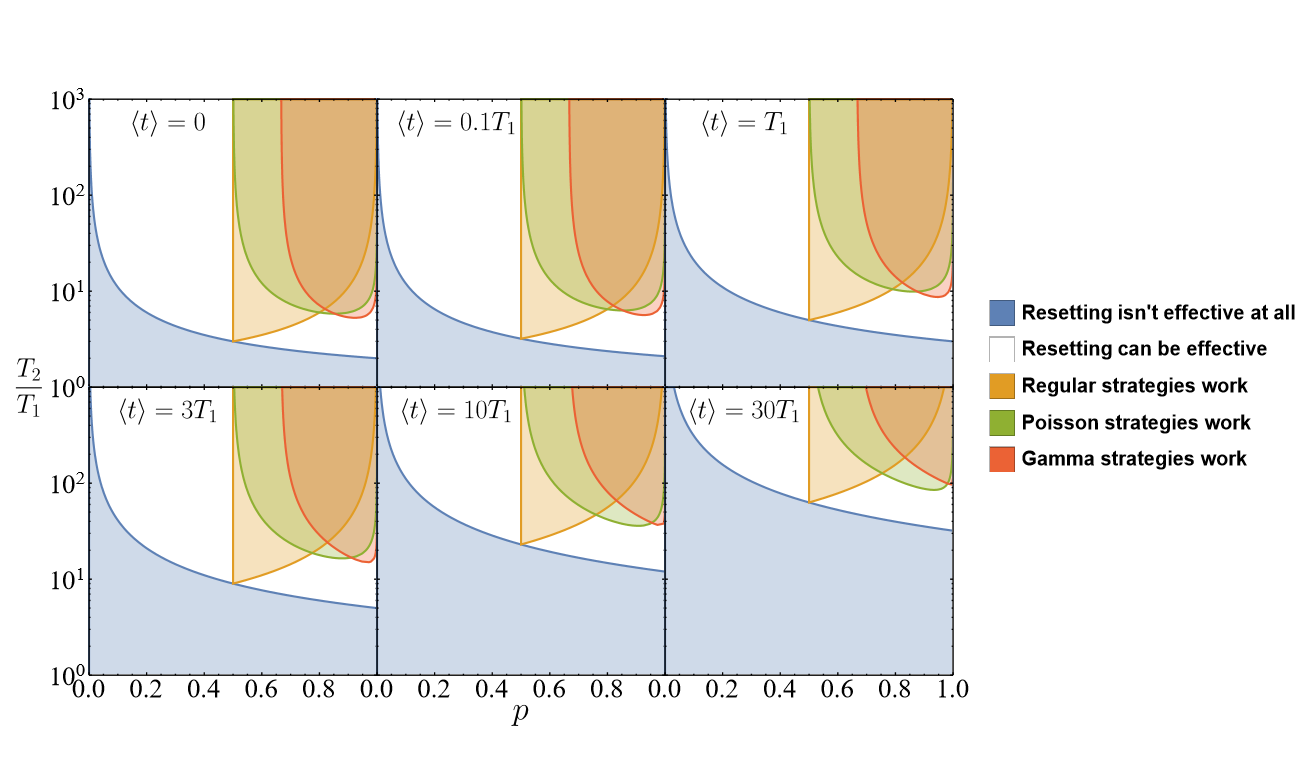}
        \caption{A diagram in the plane of dimensionless parameters $p$ and $T_2/T_1$ representing applicability regions of the restart protocols listed in Table \ref{Table1} for the random process with completion time probability density ${P(T) = p \delta(T - T_1) + (1-p) \delta(T - T_2)}$. Different subplots correspond to different values of the time penalty $\langle t \rangle$.
        }
        \label{fig:applicability_mean_delta}
    \end{figure}
\end{center} 
\end{widetext}

\begin{widetext}
\begin{table*}
\begin{center}
\begin{tabular}{|l|l|l|l|l|l|l|l|}
\hline
Protocol   & Applicability condition &  Range for effective $\tau$, $r$ or $\beta$ & Recommended period/rate & Efficiency  \\ \hline
Regular1        &   $m+\langle t\rangle< \langle |T-m|\rangle$    &  $m\le\tau< \langle |T-m|\rangle-\langle t\rangle$  & $\tau_{0}=m$     &  $\eta_1= \frac{\langle |T-m|\rangle - m - \langle t\rangle}{\langle T \rangle + 
 \langle t \rangle}$       \\
  &       &      &   &    \\
Regular2             & $m<\frac12\langle T\rangle-\frac12\langle t\rangle$ &  $m\le\tau< \frac12\langle T\rangle-\frac12\langle t\rangle$  &  $\tau_{0}=m$    &  $\eta_2 \ge 1 - \frac{2 (m + \langle t \rangle)}{ \langle T \rangle + \langle t \rangle}$         \\
  &       &       &  &  \\
Poisson1       & $\langle T^2 \rangle \ge2 \langle T \rangle  (\langle T\rangle+ \langle t \rangle)$  & $0 < r < \frac{\langle T^2 \rangle (\langle T^2 \rangle -2 \langle T \rangle  (\langle T\rangle+ \langle t \rangle))}{(\langle T\rangle+\langle t \rangle)(\langle T \rangle \langle T^3 \rangle-\langle T^2 \rangle^2)}$     & see Eq. (\ref{eq:r^m_opt_main})          & see Eq. (\ref{eq:eta_3})          \\
   &       &       &  &  \\
Poisson2     & $\langle T^2 \rangle \ge2 \langle T \rangle  (\langle T\rangle+ \langle t \rangle)$  & $0<r< \frac{3(\langle T^2 \rangle - 2 \langle T \rangle(\langle T\rangle+ \langle t\rangle))}{\langle T^3 \rangle}$     & see Eq. (\ref{eq:r^c_opt_main})          & see Eq. (\ref{eq:eta_4})          \\
   &     &    &      &  \\
Poisson3          &  $\langle T^2 \rangle \ge2 \langle T \rangle  (\langle T\rangle+ \langle t \rangle)$   & $ 0<r<\frac{\langle T^2 \rangle \left[\langle T^2 \rangle-2\langle T \rangle (\langle t\rangle+\langle T\rangle)\right]}{(\langle t\rangle+\langle T\rangle)  \langle T \rangle \langle T^3 \rangle}$   & see Eq. (\ref{eq:r^ms_opt_main})         & see Eq. (\ref{eq:eta_5})           \\
  &       &    &      &  \\
Gamma          &   $\langle T^{3}\rangle\ge 3 (\langle T\rangle+\langle t\rangle)\langle T^2\rangle$ & $0<\beta<\frac{\langle T^3\rangle-3(\langle T\rangle + \langle t\rangle) \langle T^2\rangle}{(\langle T\rangle +\langle t\rangle)\langle T^3\rangle+\langle T^4\rangle}$     & numerically available         & numerically available           \\
\hline
\end{tabular}
\end{center}
\caption{The table contains various restart protocols that are guaranteed to reduce the average completion time of the process. For each strategy, the table’s columns specify the following: the condition under which the corresponding strategy is applicable; the resulting range for an effective period (for periodic protocols) or rate parameter (for Poisson or Gamma protocols); the optimal period or rate parameter providing the maximum guaranteed effectiveness for the corresponding protocol; an estimate for the resulting maximum guaranteed efficiency.}
\label{Table1}
\end{table*}
\end{widetext}



    \begin{widetext}
\begin{center}
    \begin{figure}
        \includegraphics[width=0.7\textwidth]{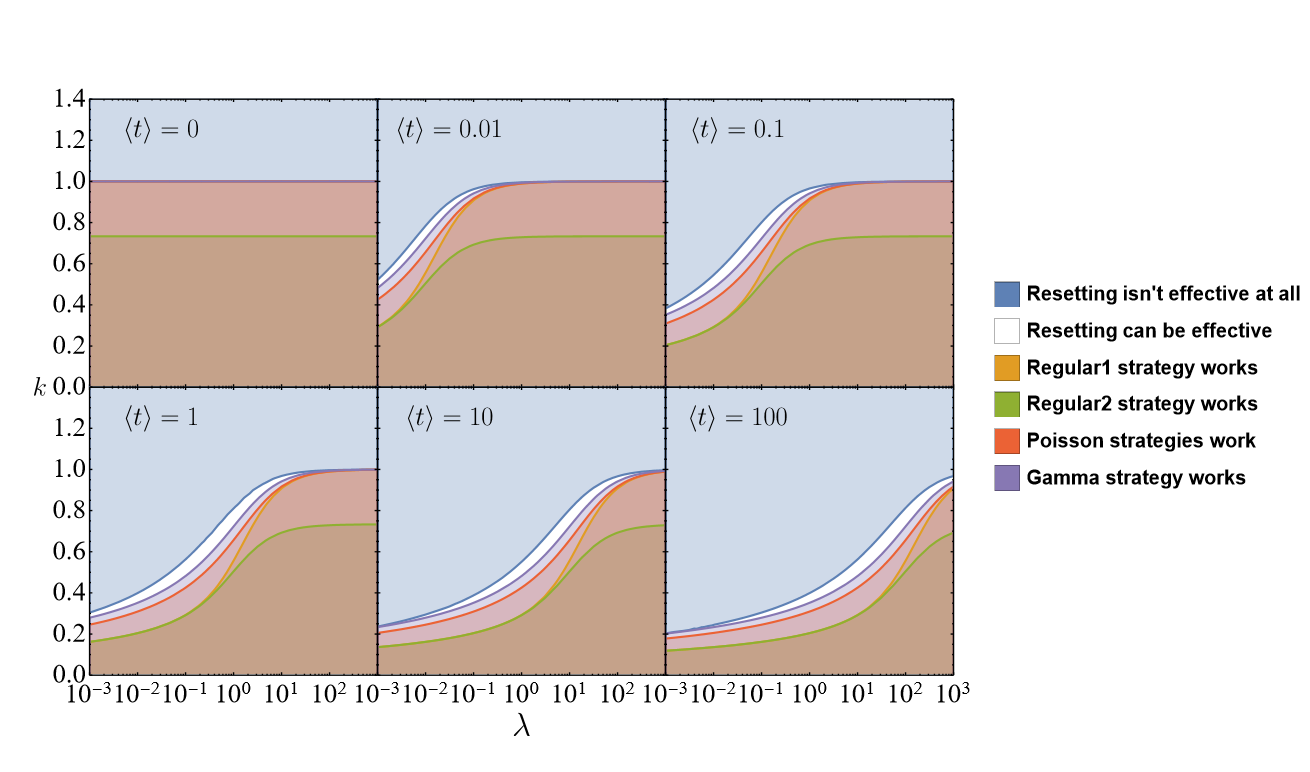}
        \caption{A diagram in the plane of dimensionless parameters $p$ and $T_2/T_1$ representing applicability regions of the restart protocols listed in Table \ref{Table1} for the random process with Weibull probability density $P(T) = \frac{k}{\lambda}(\frac{T}{\lambda})^{k-1}e^{-(\frac{T}{\lambda})^k}$ of completion time. Different subplots correspond to different values of the time penalty  $\langle t \rangle$. The data were generated via numerical analysis of Eq. (\ref{completion_time}) and applicability conditions presented in Table \ref{Table1}. 
        }
        \label{fig:applicability_mean_Weibull}
    \end{figure}
\end{center} 
\end{widetext}

  \begin{widetext}
\begin{center}
    \begin{figure}
        \includegraphics[width=0.7\textwidth]{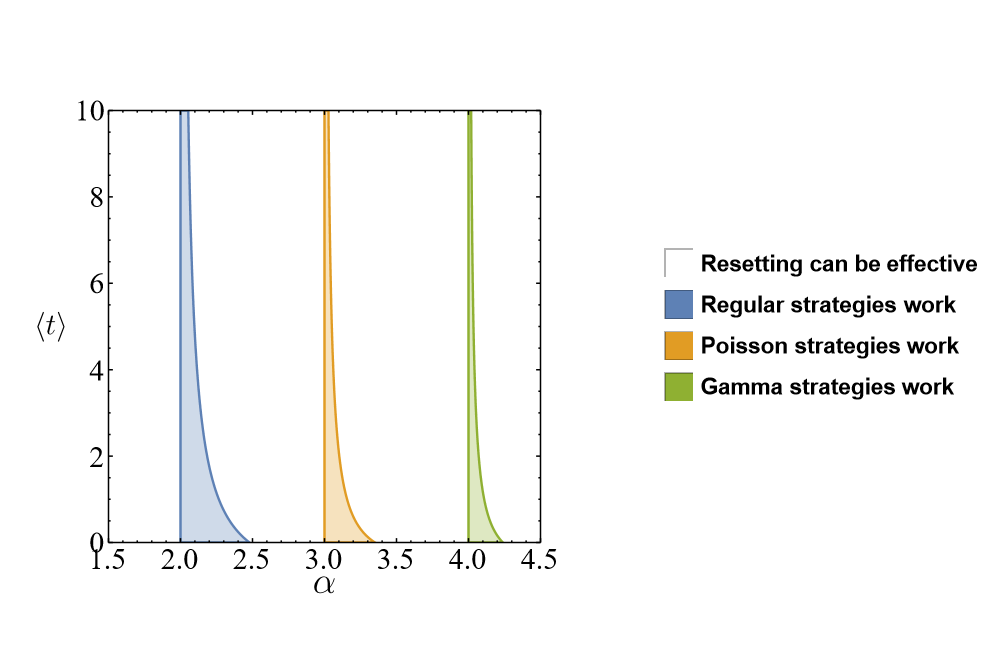}
        \caption{A diagram in the plane of parameters $\alpha$ and $\langle t\rangle$ representing applicability regions of the restart protocols listed in Table \ref{Table1} for the discrete power-law  probability density $P(T) = \frac{1}{\zeta(\alpha)}\sum_{k=1}^{\infty}\frac{\delta(\frac{T}{T_0}-k)}{(\frac{T}{T_0})^\alpha}$ of completion time, where $\zeta(\alpha)$ is the Riemann zeta function. Different subplots correspond to different values of the time penalty  $\langle t \rangle$. The data were generated via numerical analysis of Eq. (\ref{completion_time}) and the applicability conditions presented in Table \ref{Table1}. 
        }
        \label{fig:applicability_mean_Power}
    \end{figure}
\end{center} 
\end{widetext}

\section{Optimization of the success probability}
So far, we have been talking about optimizing the average completion time of a stochastic process.
Meanwhile, restart can also be used to increase the probability of observing a desired outcome of a random process with several alternative completion scenarios \cite{Belan_PRL_2018}. Examples of such processes include random search with multiple targets \cite{Condamin_2007a, Condamin_2007b, Condamin_2008, Meyer_2012, Calandre_2012, Calandre_2014, Benichou_2015, Cencetti_2016, Krapivsky_2017, Dobramysl_2018,Pal_2019a}, random search with mortality \cite{Lohmar_2009, Abad_2012, Abad_2013a, Yuste_2013, Abad_2013b, Abad_2015, Campos_2015, Meerson_2015, Grebenkov_2017}, chemical reactions with competing paths \cite{Rehbein_2011,Rehbein_2015,Martin-Soomer_2016}, folding of biopolymer into one of several native states \cite{Solomatin_2010,Marek_2011,Hyeon_2012,Paudel_2014,Hinczewski_2016,Pierse_2017}.
In this section, we provide constructive criteria of when restart increases the chances that a random process ends in the desired way. 

As a model consider a random process with two completion scenarios - "success"\ and "failure". 
The process is characterized by a random completion time $T$ having a probability density of $P(T)$. The latter can be represented as the superposition $P(T)=P^s(T)+P^f(T)$, where $P^s(T)$ and $P^f(T)$ denote the contribution of successful and unsuccessful trials, respectively. Note that the normalization of the function $P^s(T)$ determines the `undisturbed' probability of success of $p$: $p=\int_0^\infty P^s(T)dT$.

As previously, the restart protocol ${\cal R}$ is determined by a sequence of time intervals $\tau_1,\tau_2,\dots$ which specifies the restart moments. 
We will say that the protocol is effective if its implementation increases the probability of success,  i.e. $p_{{\cal R}}>p$. The corresponding restart efficiency is defined as
\begin{eqnarray}
    \chi=\frac{p_{{\cal R}}-p}{1-p}. 
\end{eqnarray}
For useful protocols, we have $0<\chi\le 1$.

If the process is restarted in strictly regular fashion with period $\tau$, then the resulting probability of observing a successful outcome is equal to \cite{Belan_PRL_2018}
\begin{eqnarray}
\label{p_r_reg}
    p_\tau=\frac{\int_0^\tau P^s(T)dT}{\int_0^\tau P(T)dT}.
\end{eqnarray}

The probability of success $p_r$ for the process under Poisson restart at rate $r$ has the following form \cite{Belan_PRL_2018}
\begin{equation}
\label{p_r_exp}
p_r=\frac{\tilde P^s(r)}{\tilde P(r)},
\end{equation}
where $\tilde P^s(r)$ and $\tilde P(r)$ denote the Laplace transforms of, respectively, $P^s(T)$ and $P(T)$ evaluated at $r$.

Finally, for the gamma protocol with rate parameter $\beta$ and shape parameter $k=1$, the resulting success probability is given by the following expression (see Appendix \ref{appendix4})
\begin{eqnarray}
\label{p_r_gamma}
    p_\beta=\frac{\beta\partial_\beta \tilde P^s(\beta)-\tilde P^s(\beta)}{\beta\partial_\beta \tilde P(\beta)-\tilde P(\beta)}.
\end{eqnarray}

Note that Eqs. (\ref{p_r_reg}-\ref{p_r_gamma}) are valid in the presence of an arbitrarily distributed random penalty for restart as long as the penalty is uncorrelated with the outcome \cite{Belan_PRL_2018}.

\subsection{Regular strategy}
Suppose the condition
\begin{eqnarray}
\label{p_sc_for_reg}
 m_s<m,   
\end{eqnarray}
is met, where $m_s$ is the median completion time of successful attempts and $m$ is the unconditional median completion time of the stochastic process of interest. These metrics are defined by the relations $p\int_0^{m_s}dTP^s(T)=1/2$ and $(1-p)\int_0^{m_f}dTP^f(T)=1/2$. Then, from (\ref{p_r_reg})  it is easy to see that the restart with a period $\tau$ belongs to the interval
\begin{eqnarray}
\label{p_r_reg1}
m_s<\tau<m,  
\end{eqnarray}
is effective for sure, since in this case the following estimate is valid
\begin{eqnarray}
p_\tau=\frac{{p}+2\int_{m_s}^\tau dTP^s(T)}{{1}-2\int_\tau^{m}dTP(T)}>p.
\end{eqnarray}

Unfortunately, the optimal period providing the greatest guaranteed efficiency $\chi$ on the interval (\ref{p_r_reg1}) cannot be expressed in terms of $m_s$ and $m$ but depends on the fine details of the probability density $P(T)$.

\subsection{Poisson strategy}
A simple sufficient condition for the existence of an effective Poisson restart protocol found in \cite{Belan_PRL_2018} reads  
\begin{eqnarray}
\label{p_r_pois_sc}
\langle T_s\rangle<\langle T\rangle,
\end{eqnarray} 
where $\langle T_s\rangle=p^{-1}\int_0^\infty dTP^s(T)T$ is the average completion time of successful trials, while $\langle T\rangle$ is the unconditional mean completion time. This criterion, however, does not say anything about how to choose an efficient restart rate since knowledge of linear statistical moments $\langle T_s\rangle$ and $\langle T\rangle$ alone is not enough for these purposes. Below, we show that adding information about the second moment of the random completion time $\langle T^2\rangle$ allows us to formulate a constructive criterion for the effectiveness of a Poisson restart. 

The probability of success given by Eq. (\ref{p_r_exp}) can be estimated from below as
\begin{eqnarray}
\label{p_r_estimate1}
    p_r \ge \frac{(1 - r\langle T_s \rangle)(\langle T^2\rangle r + \langle T\rangle)}{(\langle T^2\rangle-\langle T\rangle^2)r +\langle T
\rangle}p,
\end{eqnarray}
where we used the bound $P^s(r)\ge p\left(1 - r\langle T_s\rangle\right)$, which directly follows from (\ref{eq:P_r_mod_main}), and the inequality \cite{Zubkov_1999} (also see Appendix \ref{appendix3}) 
\begin{equation}
\label{eq:P_r_mod_le_main}
 P(r) \le 1 - \frac{r\langle T\rangle^2}{r\langle T^2\rangle + \langle T\rangle}.
\end{equation}

We see from (\ref{p_r_estimate1}) that if the existence condition (\ref{p_r_pois_sc}) is met, then a Poisson restart with a rate enclosed inside the interval 
\begin{eqnarray}
\label{answer_for_range}
    0<r<\frac{\langle T \rangle \left(\langle T \rangle - \langle T_s \rangle\right)}{\langle T^2 \rangle \langle T_s \rangle},
\end{eqnarray}
increases the chances of success, i.e. $p_r>p$. As can be found by examining the right side of the inequality (\ref{p_r_estimate}) for an extremum, the point $r_0$ given by Eq. (\ref{p_r_range}) in Appendix \ref{appendix5}
provides the maximum guaranteed gain for the given values of $\langle T_s\rangle$, $\langle T\rangle$, and $\langle T^2\rangle$ provided the use of the estimates described above. 
The resulting efficiency is estimated from below, as shown in Eq. (\ref{chi_2}).

Let us demonstrate another road to the range of efficient rates. Alternatively, using the results of \cite{Brook_1966}, we can evaluate the Laplace transform from above as
\begin{eqnarray}
    \tilde P(r)\le 1-\frac{\langle T\rangle^2}{\langle T^2\rangle}+\frac{\langle T\rangle^2}{\langle T^2\rangle} e^{-\frac{\langle T^2\rangle}{\langle T\rangle}r},
\end{eqnarray}
and then from Eq. (\ref{p_r_exp}) one gets
\begin{eqnarray}
\label{p_r_estimate}
    p_r \ge \frac{\langle T^2\rangle(1 - r\langle T_s \rangle)}{\langle T^2\rangle-\langle T\rangle^2+   \langle T\rangle^2 e^{-\frac{\langle T^2\rangle}{\langle T\rangle}r}}p.
\end{eqnarray}
By requiring the expression on the right side of the last inequality to exceed the undisturbed probability of success $p$, we find that if the condition $\langle T_s\rangle<\langle T\rangle$ is satisfied, then all rates belonging to the interval $0<r<r_c$, where $r_c$ represents the solution of the transcendental equation $1-\frac{\langle T^2\rangle\langle T_s\rangle}{\langle T\rangle^2}r-e^{-\frac{\langle T^2\rangle}{\langle T\rangle}r}=0$, increase the chances of success. 
Interestingly, at $\langle T\rangle-\langle T_s\rangle\ll\langle T\rangle$, one obtains the boundary $r_c\approx 2\frac{\langle T\rangle-\langle T_s\rangle}{\langle T^2\rangle}$, which twice exceeds the corresponding value dictated by Eq. (\ref{answer_for_range}).

\subsection{Gamma strategy}
Finally, let us analyze the case of the gamma protocol.
From (\ref{mean_T_Gamma}), we see that if the condition 
\begin{eqnarray}
\label{p_r_gamma_sc}
    \langle T^{2}\rangle\ge \langle T^{2}_s\rangle,
\end{eqnarray} 
is met, 
then $\partial_\beta p_\beta>0$ at $\beta=0$ and, therefore, the inequality (\ref{p_r_gamma_sc}) represents a sufficient condition for the existence of an efficient gamma strategy.
Note, however, that 
the fixed pair $\langle T^2\rangle$ and $\langle T^2_s\rangle$ determines only the slope of $p_\beta$ in its dependence of $\beta$ at the point $\beta=0$, without saying anything about its behavior for non-zero $\beta$. 
Let us show that efficient rates can be specified if the third-order moment $\langle T^3\rangle$ is additionally known.

\begin{widetext}
\begin{table*}
\begin{center}
\begin{tabular}{|l|l|l|l|l|l|l|l|}
\hline
Protocol    & Applicability condition &  Range for effective period/rate & Recommended period/rate & Efficiency  \\ \hline
Regular          &   $m_s<m$   &  $m_s<\tau<m$ & -     &  -      \\
  &        &      &   &    \\
Poisson 1               & $\langle T_s\rangle<\langle T\rangle$ &  $0<r<\frac{\langle T \rangle \left(\langle T \rangle - \langle T_s \rangle\right)}{\langle T^2 \rangle \langle T_s \rangle}$  &  see Eq. (\ref{p_r_range})   &  see Eq. (\ref{chi_2})         \\
 &       &       &  &  \\
Poisson 2     &  $\langle T_s\rangle<\langle T\rangle$  &  numerically available    &   numerically available        & numerically available          \\
  &       &       &  &  \\
Gamma 1      & $\langle T_s^2\rangle<\langle T^2\rangle$  & $0<\beta<\frac{3(\langle T^2\rangle-\langle T_s^2\rangle)}{3\langle T^3\rangle+\langle T_s^3\rangle}$ & numerically available        & numerically available          \\
  &      &       &  &  \\
Gamma 2       & $2\langle T_s^2\rangle<\langle T^2\rangle$  & $0<\beta<\frac{\langle T^2 \rangle - 2 \langle T_s^2 \rangle}{\langle T^3 \rangle}$    & numerically available        & numerically available          \\
  &      &       &  &  \\
Gamma 3       & $2\langle T_s^2\rangle<\langle T^2\rangle$  & $0<\beta<\frac{\langle T^2 \rangle (\langle T^2 \rangle - 2 \langle T_s^2 \rangle)}{\langle T^3 \rangle (\langle T^2 \rangle + 2 \langle T^2_s \rangle)}$    & numerically available        & numerically available          \\
\hline
\end{tabular}
\end{center}
\caption{A table summarizes various restart protocols guaranteed to increase the probability of success. For each strategy, the table’s columns specify the following: the condition when the corresponding strategy works; the resulting range for an effective period (for periodic protocols) or rate (for Poisson or Gamma protocols); the optimal period or rate providing the maximum guaranteed effectiveness; an estimate for the resulting maximum guaranteed efficiency.}
\label{Table2}
\end{table*}
\end{widetext}

The success probability dictated by (\ref{p_r_gamma}) can be estimated from below in several different ways.
Namely, first let us exploit inequality (\ref{eq:P_r_classic_main}) at $l = 1$ and $l = 2$ for $\beta\partial_\beta\tilde P(\beta)$ and $\tilde{P}(\beta)$, respectively. 
Next, let us estimate both terms in the denominator of Eq. (\ref{p_r_gamma}) via the inequality \cite{Zubkov_1999}
\begin{eqnarray}
    \label{eq:upper_estimation}
    \tilde{P}(\beta) \le \sum_{k=0}^{2l} (-1)^k \frac{\beta^k \langle T^k \rangle}{k!},
\end{eqnarray}
at $l = 1$. 
This yields
\begin{eqnarray}
    p_\beta\ge\frac{6-3\langle T_s^2\rangle \beta^2- \langle T_s^3\rangle \beta^3}{3(2-\langle T^2\rangle \beta^2+\langle T^3\rangle \beta^3)}p,
\end{eqnarray}
where we assumed $1-\frac{\beta^2}{2}\langle T^2\rangle+\frac{\beta^3}{2}\langle T^3\rangle\ge 0$. 
Then, requiring that the corresponding bound is greater than the undisturbed probability of success $p$, we obtain the interval
\begin{eqnarray}
    0<\beta<\frac{3(\langle T^2\rangle-\langle T_s^2\rangle)}{3\langle T^3\rangle+\langle T_s^3\rangle},
\end{eqnarray}
for rates that guarantee to increase the chances of success. The interval is non-empty as long as the condition (\ref{p_r_gamma_sc}) is met.


Alternatively, one can use (\ref{eq:P_r_classic_main}) with $l = 1$ to bound $\tilde{P}(\beta)$ in the numerator of (\ref{p_r_gamma}), while leaving other estimates unchanged.
This gives 
\begin{eqnarray}
\label{p_r_gamma_bound_1}
    p_\beta\ge \frac{2 - 2\beta^2 \langle T_s^2 \rangle}{2 - \beta^2 \langle T^2 \rangle + \beta^3 \langle T^3 \rangle}p.
\end{eqnarray}
The same line of reasoning as described above gives the following interval of efficient rates
\begin{eqnarray}
    0<\beta<\frac{\langle T^2 \rangle - 2 \langle T_s^2 \rangle}{\langle T^3 \rangle}.
\end{eqnarray}
In contrast to the previous case, here we face a stronger applicability condition $\langle T^{2}\rangle\ge 2\langle T^{2}_s\rangle$.

A Similar result can be obtained if we estimate both terms entering the numerator of expression for $p_\beta$ using (\ref{eq:P_r_classic_main}) with $l = 1$ and exploit (\ref{eq:P_r_mod_le_main}) to estimate the term $-\partial_\beta \tilde{P}(\beta) = \langle T \rangle \tilde{Q}(\beta)$ in the denominator.
For the probability of success we then find 
\begin{eqnarray}
\label{p_r_gamma_bound_2}
    p_\beta\ge \frac{(2 - 2\beta^2 \langle T_s^2 \rangle)(\langle T^2 \rangle + \beta \langle T^3 \rangle)}{2 \langle T^2 \rangle - \beta^2 \langle T^2 \rangle^2 + \beta^3 \langle T^3 \rangle \langle T^2 \rangle + 2 \beta \langle T^3 \rangle}p.
\end{eqnarray}
This estimate yields a narrower interval of effective rates
\begin{eqnarray}
    {0<\beta<\frac{\langle T^2 \rangle (\langle T^2 \rangle - 2 \langle T_s^2 \rangle)}{\langle T^3 \rangle (\langle T^2 \rangle + 2 \langle T^2_s \rangle)}},
\end{eqnarray} 
with the previous condition of applicability ${\langle T^2\rangle>2\langle T_s^2\rangle}$. 

To determine the best rates that provide the maximum guaranteed efficiency of gamma restart at the intervals described above and to calculate the resulting efficiencies, one needs to solve high-order algebraic equations, which is more convenient to do by numerical methods.
    


\begin{widetext}
\begin{center}
    \begin{figure}[t]
        \includegraphics[width=0.7\textwidth]{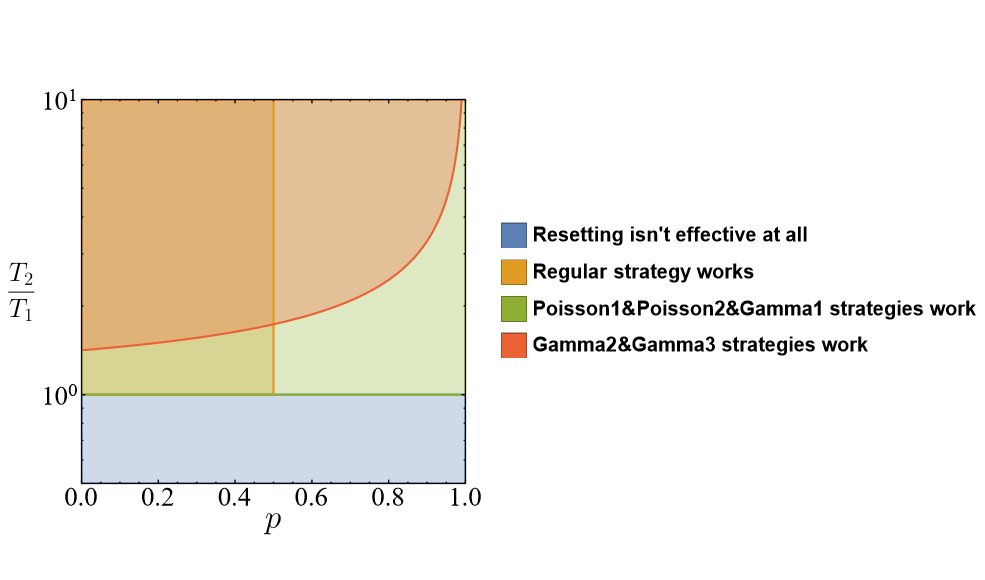}
        \caption{A diagram in the plane of dimensionless parameters $p$ and $T_2/T_1$ representing applicability regions of the restart protocols listed in Table \ref{Table2} for the random process with completion time probability density $P(T)=P^s(T)+P^f(T)$ with $P^s(T)=p \delta(T - T_1)$ and $P^f(T)=(1-p) \delta(T - T_2)$.
        }
        \label{fig:applicability_prob_delta}
    \end{figure}
\end{center} 
\end{widetext}

\section{Discussion and conclusion}
As explained in the introduction, the evaluation of the optimal restart period $\tau_\ast$ requires the knowledge of the exact completion time probability density $P(T)$.
Trying to address this issue, the authors of \cite{Luby_1993} have developed a restart protocol, nowadays known as the Luby strategy, which has the following remarkable property:  regardless of the statistical details of the original process, the average completion time of this process subject to the Luby strategy exceeds the expectation of completion time provided by the optimal periodic protocol by no more than a logarithmic factor.

While ingenious and elegant, Luby's strategy suffers from two serious drawbacks. First, it only applies to processes with discrete completion times. In the case of continuous time, as was recently shown by Lorentz \cite{Lorenz_2021}, such a universal strategy simply does not exist. 
Secondly, and even more importantly, even for discrete-time processes, by applying the Luby strategy, we can only be sure that the resulting mean completion time will not be too bad compared to the optimal value achieved by the best periodic strategy.
In other words, there is no guarantee that the Luby strategy will not degrade performance as compared to the restart-free case \cite{Lorenz_2021}.
Thus, in the complete absence of information about the statistical properties of the process, the only way to insure from decrease of performance is not to restart the process at all.



Significant progress in solving the problem of choosing an effective restart policy was achieved in work \cite{Eliazar_JPA_2021}, whose authors formulated relatively simple constructive criteria for the effectiveness of periodic restarts for random processes with partially specified statistical information.
Overcoming the limitations of the previously known existence results \cite{Reuveni_PRL_2016,starkov2023universal}, these criteria offer a specific restart period that is guaranteed to reduce the average completion time of the random process.
Motivated by this progress, in this paper, we generalized one of the criteria proposed in \cite{Eliazar_JPA_2021} to the case of a non-zero time penalty for restart and also constructed several new criteria, some of which concern the case of stochastic restart. 
In addition, we have offered the first examples of the constructive criterion of restart efficiency in the context of success probability optimization.
The results of the analysis are summarized in tables (\ref{Table1}) and (\ref{Table2}).

For the sake of illustration, we 
analyzed the efficiency of various constructive  criteria using the weighted sum of two delta-functions ${P(T) = p \delta(T - T_1) + (1-p) \delta(T - T_2)}$ as a model distribution.
Figure \ref{fig:applicability_mean_delta} represents the diagrams obtained by analyzing the applicability conditions given by Eqs. (\ref{eq6}),(\ref{eq10}), (\ref{T_r_pois_sc}) and (\ref{gamma_sc}) in terms of the dimensionless parameters $p$ and $T_1/T_2$ for different values of time penalty $\langle t\rangle$.
The Blue area corresponds to the case when no restart strategy can be beneficial. For the region with the orange boundary  the regular strategies described above are applicable. The area above the green line corresponds to the scenario when Poisson strategies work. Next, our gamma strategies work in the region of the diagram bounded by the red line. Finally, in the white parts of the diagram, restart is potentially beneficial, but our criteria fail to capture that.
Similarly, we analyzed the Weibull completion time distribution $P(T) = \frac{k}{\lambda}(\frac{T}{\lambda})^{k-1}e^{-(\frac{T}{\lambda})^k}$ and discrete power-law distribution $P(T) = \frac{1}{\zeta(\alpha)}\sum_{k=1}^{\infty}\frac{\delta(\frac{T}{T_0}-k)}{(\frac{T}{T_0})^\alpha}$, see Figs. \ref{fig:applicability_mean_Weibull} and \ref{fig:applicability_mean_Power}, respectively. 
These examples demonstrate that all the criteria formulated here represent sufficient but not necessary conditions. 
Indeed, as illustrated by the white areas in Fig. \ref{fig:applicability_mean_delta} and Fig. \ref{fig:applicability_mean_Power}, there exist random processes for which none of the applicability conditions given by Eqs. (\ref{eq6}), (\ref{eq10}), (\ref{T_r_pois_sc}) and (\ref{gamma_sc}) are fulfilled, but an effective strategy still exists. 
Note also that there are regions where only stochastic protocols work. 
This justifies our motivation to construct effective Poisson and Gamma strategies.

Also, we used the toy model $P(T)=P^s(T)+P^f(T)$ with $P^s(T)=p \delta(T - T_1)$ and $P^f(T)=(1-p) \delta(T - T_2)$ to illustrate the efficiency of different restart protocols in the context of success probability optimization.
The diagrams represented in Figure (\ref{fig:applicability_prob_delta}) show the regions where conditions (\ref{p_sc_for_reg}), (\ref{p_r_pois_sc}), and (\ref{p_r_gamma_sc}) are satisfied.

As discussed above, one can construct an infinite number of criteria for Poisson and Gamma protocols by adding statistical moments of larger order into estimates defined by Eqs. (\ref{eq:P_r_classic_main}), (\ref{eq:der_P_r}) and (\ref{eq:upper_estimation}).
Here we restricted ourselves to the simplest possible criteria involving the minimal number of low-order moments required to specify the effective values of rate parameters.
In particular, we showed that one needs to know at least three moments $\langle T\rangle$, $\langle T^2\rangle$, and $\langle T^3\rangle$ in the case of the Poisson restart.
At the same time, as discussed in Section II, even for fixed values of  $\langle T\rangle$, $\langle T^2\rangle$, and $\langle T^3\rangle$, the interval of effective rates is not unique: using different estimates for the Laplace transform of the completion time probability density $P(T)$ one can construct a range of efficient strategies with different recommended control parameters (see Table I).

Since the knowledge of all statistical moments would allow to recover the probability density $P(T)$, and thus to find the best possible control parameter ($r_\ast$ or $\beta_\ast$), one may expect that the recommended rate $r_0$ ($\beta_0$) tends to the true optimal value $r_\ast$ ($\beta_\ast$) with the growth of the number of moments $l$ used in estimates (\ref{eq:P_r_classic_main}), (\ref{eq:der_P_r}) and (\ref{eq:upper_estimation}).
If this is so, then the next question naturally arises: how quickly does convergence occur?
Although the answer may be non-universal (i.e., $P(T)$-dependent), we believe that this issue represents an interesting avenue for future research.



\subsection{Acknowledgements}

The work was supported by the Russian Science Foundation (RSF), project $\#$22-72-10052.
The authors are grateful to V.V. Lebedev for his useful comments. I.S. Nikitin would like to thank A.M. Zubkov for referring to relevant literature on statistical inequalities.

\begin{widetext}

\appendix

\section{Derivation of Eq. (\ref{completion_time})}
\label{appendix1}

Here we generalize the derivation proposed in \cite{Eliazar_JPA_2021} to the case of the non-zero time penalty $\langle T_{on} \rangle$.

The mean completion time $\langle T_\tau \rangle$ in the presence of periodic restart with period $\tau$ is given by 
\begin{equation}
\label{ap1eq0}
    \langle T_\tau \rangle =  \langle T_{on} \rangle + \langle \min (T, \tau) \rangle + \langle T_{\tau}' \cdot I(T \ge \tau) \rangle,
\end{equation}
where $T_\tau'$ is the statistically independent copy of $T_\tau$ and $I(...)$ denotes the indicator random variable. 
The last term in the left-hand side can be transformed using that $T_{\tau}'$ is statistically independent on   
$T_{\tau}$:
\begin{equation}
\label{ap1eq1}
    \langle T_{\tau}' \cdot I(T \ge \tau) \rangle = \langle T_{\tau}' \rangle \langle I(T \ge \tau) \rangle = \langle T_{\tau}' \rangle \int_\tau^\infty P(T) dT = \langle T_{\tau} \rangle \left(1 - \int_0^\tau P(T) dT\right).
\end{equation}
Next, the term 
$\langle \min (T, \tau) \rangle$ can be rewritten based on the identity 
\begin{equation}
    2 \min(T, \tau) = T + \tau - |T - \tau|.
\end{equation}
Therefore 

\begin{equation}
\label{ap1eq2}
     \langle \min(T, \tau) \rangle = \frac12(\langle T \rangle + \tau - \langle |T - \tau| \rangle).
\end{equation}
Substituting Eqs. (\ref{ap1eq1}) and (\ref{ap1eq2}) into Eq. (\ref{ap1eq0}), we find Eq. (\ref{completion_time}).


\section{Derivation of Eq. (\ref{mean_T_Gamma})}
\label{appendix2}

Assume that a stochastic process is subject to the stochastic restart protocol, where random intervals between restarts are independently sampled from the Gamma distribution $\rho(\tau)=\frac{\beta^k}{\Gamma(\beta)}\tau^{k-1}e^{-\beta \tau}$ with rate parameter $\beta$ and natural shape parameter $k$.
Then the random completion time $T_{\beta}$ obeys the following renewal equation
\begin{eqnarray}
\label{chain11111}
&&T_{\beta}=T_{on}+TI(T< \tau)+(\tau+T_{\beta}') I(T\ge \tau),
\end{eqnarray}
where $T_{\beta}'$ is a statistically independent replica of $T_{\beta}$.
Let us average this relation over the statistics of the original process and of the inter-restart intervals.
This gives
\begin{equation}
\langle T_{\beta}\rangle= \langle T_{on}\rangle+\langle TI(T< \tau)\rangle+ \langle\tau I(T\ge  \tau)\rangle+
\langle T_{\beta}'  I(T\ge \tau)\rangle.
\end{equation}
Since $\langle TI(T< \tau)\rangle=\int_0^\infty dT P(T)T  \int_T^\infty d\tau \rho(\tau) $, $ \langle\tau I(T\ge  \tau)\rangle=\int_0^\infty dT P(T)  \int_0^T d\tau \rho(\tau)\tau$, $\langle T_{\beta}'  I(T\ge \tau)\rangle=\langle T_{\beta}' \rangle\langle I(T\ge  \tau)\rangle$, $1-\langle I(T\ge \tau)\rangle=\int_0^\infty dT P(T)  \int_T^\infty d\tau \rho(\tau)$ and $\langle T_{\beta}' \rangle=\langle T_{\beta}\rangle$, we find the closed-form expression for the expected completion time 
\begin{equation}
\label{stochastic}
\langle T_{\beta}\rangle=\frac{\langle T_{on}\rangle+\int_0^\infty dT P(T)T  \int_T^\infty d\tau \rho(\tau) +\int_0^\infty dT P(T)  \int_0^T d\tau \rho(\tau)\tau}{\int_0^\infty dT P(T)  \int_T^\infty d\tau \rho(\tau)}.
\end{equation}
Next we obtain
\begin{eqnarray}
&&\int_0^\infty dT P(T)T  \int_T^\infty d\tau \rho(\tau)=\frac{(-1)^{k-1}\langle T\rangle \beta^k}{\Gamma(k)}\frac{d^{k-1}}{d\beta^{k-1}}\int_0^\infty dT Q(T)  \int_T^\infty d\tau e^{-\beta\tau}=\frac{(-1)^{k-1}\langle T\rangle \beta^k}{\Gamma(k)}\frac{d^{k-1}}{d\beta^{k-1}} \frac{\tilde Q(\beta)}{\beta}=\\
.&&=\frac{(-1)^{k} \beta^k}{\Gamma(k)}\frac{d^{k-1}}{d\beta^{k-1}}\left[\frac{1}{\beta}\frac{d\tilde P(\beta)}{d\beta}\right],\\
&&\int_0^\infty dT P(T)  \int_0^T d\tau \rho(\tau)\tau=\frac{(-1)^{k-1} \beta^k}{\Gamma(k)}\frac{d^{k-1}}{d\beta^{k-1}}\int_0^\infty dT P(T)  \int_0^T d\tau e^{-\beta\tau}\tau=\\
&&=\frac{(-1)^{k-1} \beta^k}{\Gamma(k)}\frac{d^{k-1}}{d\beta^{k-1}}\left[\frac{1}{\beta}\left(\frac{1}{\beta}-\frac{1}{\beta}\tilde P(\beta)+\frac{d\tilde P(\beta)}{d\beta} \right)\right],\\
&&\int_0^\infty dT P(T)  \int_T^\infty d\tau \rho(\tau)=\frac{(-1)^{k-1} \beta^k}{\Gamma(k)}\frac{d^{k-1}}{d\beta^{k-1}}\int_0^\infty dT P(T)  \int_T^\infty d\tau e^{-\beta\tau}=\frac{(-1)^{k-1} \beta^k}{\Gamma(k)}\frac{d^{k-1}}{d\beta^{k-1}} \frac{\tilde P(\beta)}{\beta},
\end{eqnarray}
where we introduced the auxiliary probability density $Q(T)=\frac{T}{\langle T\rangle}P(T)$.

For $k=1$, we find immediately Eq. (\ref{eq:T_r_mean_main}), whereas for $k=2$, one get Eq. (\ref{mean_T_Gamma}).

\section{Derivation of Eqs. (\ref{eq:P_r_mod_main}, \ref{eq:P_r_mod_le_main})}
\label{appendix3}

Here we provide a derivation of inequalities from Ref. \cite{Zubkov_1999}. The Laplace transform of the probability distribution function $P(T)$ of a non-negative random variable $T$ is defined by
\begin{equation}
    \label{eq:P_r_def}
    \tilde{P}(r) = \int_{0}^{\infty}{e^{-rT}P(T)dT}.
\end{equation}
Let $\nu$ be a number of events in a Poisson process with rate $r$ that occurred during a time interval of random duration $T$. Then $p(\nu)$ is defined by
\begin{equation}
    \label{p_nu}
    p(\nu) = \int_{0}^{\infty}\frac{(rT)^\nu}{\nu !}{e^{-rT}P(T)dT}.
\end{equation}
Let us compute $\langle \nu^n\rangle$. It can be easily done with the definition of a Stirling number of the second kind $S(n, k)$
\begin{equation}
    \nu^{n} = \sum_{k = 0}^{n} S(n, k) \nu^{[k]}, 
\end{equation}
where $\nu^{[n]}=\nu(\nu - 1)(\nu - 2)\dots(\nu - n + 1)$. Note that 
\begin{equation}\label{d_f_s}
    \langle \nu^{[n]} \rangle = \left.\frac{d^n \langle 
 s^\nu\rangle}{ds^n}\right|_{s = 1}.
\end{equation}
Using the Eq. (\ref{p_nu}), one can find
\begin{equation}\label{f_s}
     \langle s^\nu\rangle = \sum_{\nu=0}^\infty {p(\nu) s^\nu} = \int_{0}^{\infty}\sum_{\nu=0}^\infty\frac{(rTs)^\nu}{\nu!}{e^{-rT}P(T)dT} = \int_{0}^{\infty}{e^{-(1-s)rT}P(T)dT} = \tilde{P}(r(1-s)).
\end{equation}
From Eqs. (\ref{f_s}, \ref{d_f_s}) follows
\begin{equation}\label{d_f_s1}
    \langle \nu^{[n]} \rangle =
    \left.\frac{d^n \tilde{P}(r(1-s))}{ds^n}\right|_{s = 1} = (-r)^n \left.\frac{d^n \tilde{P}(x)}{dx^n}\right|_{x = 0} = r^n \langle T^n \rangle.
\end{equation}
Then Eq. (\ref{p_nu}) and Eq. (\ref{d_f_s1}) give us the following identity
\begin{equation}\label{nu^k}
    \langle \nu^{n} \rangle = \sum_{k = 0}^{n} S(n, k) \langle \nu^{[k]} \rangle = \sum_{k = 0}^{n} S(n, k)r^k \langle T^{k} \rangle.
\end{equation}
Assume that $\langle \nu ^ 3 \rangle < \infty$, which is equivalent to $\langle T^3 \rangle < \infty$ (see Eq.(\ref{nu^k})). Consider a non-negative definite quadratic form
\begin{equation}\label{Quadratic_form}
    \langle I(\nu > 0) (\nu - 1)(z_0 + z_1 \nu)^2\rangle = \sum_{i, j = 0}^{1} z_i z_j (m_{i+j+1} - m_{i+j}),
\end{equation}
where $m_k = \langle \nu ^k I(\nu > 0)\rangle$. Its matrix is defined by
\begin{equation}
    M = 
    \begin{pmatrix}
        m_1 - Pr[\nu \ge 1] & m_2 - m_1\\
        m_2 - m_1 & m_3 - m_2
    \end{pmatrix}.
\end{equation}
Since quadratic form is non-negative, $\det(M) \ge 0$
\begin{equation}\label{Pr_nu_1}
    Pr[\nu \ge 1] \le m_1 - \frac{(m_2 - m_1)^2}{m_3-m_2}. 
\end{equation}
Using that $Pr[\nu \ge 1]=1-p(0)$ and $m_k = \langle \nu^k I(\nu > 0)\rangle = \langle \nu^k \rangle$ at $k>0$, we get 
\begin{equation}
    \label{eq:p(0) appendix 14}
    p(0) \ge 1 - \langle \nu \rangle + \frac{(\langle \nu ^2 \rangle - \langle \nu \rangle)^2}{\langle\nu^3 \rangle - \langle \nu^2 \rangle}.
\end{equation}

From Eqs. (\ref{nu^k}, \ref{eq:p(0) appendix 14}) and the identity $p(0) = \tilde{P}(r)$, one can obtain the following inequality (Eq. (\ref{eq:P_r_mod_main}) from the main text)
\begin{equation}\label{eq:P_r_ge}
    \tilde{P}(r) \ge 1 - r \langle T \rangle + \frac{r^2 \langle T^2 \rangle}{r\langle T^3 \rangle + 2 \langle T^2 \rangle}.
\end{equation}

Similarly, assuming $\langle \nu^2 \rangle < \infty$, which is equivalent to $\langle T^2 \rangle < \infty$ by Eq. (\ref{nu^k}), we can get the Eq. (\ref{eq:P_r_mod_le_main}) from the main text. Analogously to Eq. (\ref{Quadratic_form}), let us consider a non-negative definite quadratic form
\begin{equation}
    \langle I(\nu > 0)(z_0 + z_1 \nu)^2\rangle = \sum_{i, j = 0}^{1} z_i z_j m_{i+j}.
\end{equation}
Then its matrix is defined by
\begin{equation}
    M = 
    \begin{pmatrix}
        Pr[\nu > 0] & m1\\
        m_1 & m_2
    \end{pmatrix}.
\end{equation}
From the condition of non-negativity of the $\det(M)$, one can find
\begin{equation}
    Pr[\nu > 0] \ge \frac{m_1^2}{m_2}.
\end{equation}
Using $Pr[\nu > 0] = 1-p(0)$ and $m_k = \langle \nu^k I(\nu > 0)\rangle = \langle \nu^k \rangle$ at $k>0$, we get
\begin{equation}
    \label{eq:p0 appendix 43}
   p(0) \le 1 - \frac{\langle \nu \rangle ^2}{\langle \nu^2 \rangle}.
\end{equation}
From Eqs. (\ref{nu^k}, \ref{eq:p0 appendix 43}) and the identity $p(0) = \tilde{P}(r)$, one can obtain Eq. (\ref{eq:P_r_mod_le_main}) from the main text
\begin{equation}\label{eq:P_r_le}
    \tilde{P}(r) \le 1 - \frac{r\langle T \rangle ^2}{r\langle T^2 \rangle + \langle T \rangle}.
\end{equation}

\section{Derivation of Eq. (\ref{p_r_gamma})}
\label{appendix4}

As shown in Ref. \cite{Belan_PRL_2018}, the success probability in the presence of the restart is given by
\begin{eqnarray}
\label{p_R_general_eq}
    p_{{\cal R}}=\frac{\int_0^\infty d\tau \rho(\tau)\int_0^\tau dT P^s(T)}{\int_0^\infty d\tau \rho(\tau)\int_0^\tau dT P(T)}.
\end{eqnarray}
where $\rho(\tau)$ denotes the probability density of time intervals between restart events.

For Gamma distribution $\rho(\tau)=\frac{\beta^k}{\Gamma(\beta)}\tau^{k-1}e^{-\beta \tau}$,
we find from (\ref{p_R_general_eq})
\begin{eqnarray}
p_\beta=\frac{\int_0^\infty d\tau \tau^{k-1}e^{-\beta\tau}\int_0^\tau dT P^s(T)}{\int_0^\infty d\tau \tau^{k-1}e^{-\beta\tau}\int_0^\tau dT P(T)}=\frac{\partial_\beta^{k-1}\int_0^\infty d\tau e^{-\beta\tau}\int_0^\tau dT P^s(T)}{\partial_\beta^{k-1}\int_0^\infty d\tau e^{-\beta\tau}\int_0^\tau dT P(T)}=\frac{\partial_\beta^{k-1}\left(\frac{\tilde P^s(\beta)}{\beta}\right)}{\partial_\beta^{k-1} \left(\frac{\tilde P(\beta)}{\beta}\right)}.
\end{eqnarray}
For $k=2$ this yields Eq. (\ref{p_r_gamma}).

\color{black}

\section{Poisson protocols}
\label{appendix5}

\subsection{Optimization of the mean completion time}

It is easy to show that within the interval given by \eqref{effective_r_m}, the expression on the right side of the inequality \eqref{estimate_T_r_main} reaches its minimum value at the point 
\begin{equation}\label{eq:r^m_opt_main}
    r_3 = \frac{- 2 \langle T^2 \rangle {\cal T}_{on} + \sqrt{2 \langle T^2 \rangle^3\left[{\cal T}_{on} \langle T^3 \rangle-\langle T^2 \rangle^2-2 {\cal T}_{on}\langle T^2 \rangle \langle t \rangle \right]\left[\langle T^3 \rangle \langle T \rangle-\langle T^2 \rangle^2\right]^{-1}}}{{\cal T}_{on} \langle T^3 \rangle - \langle T^2 \rangle^2},
\end{equation}
where ${\cal T}_{on} = \langle T\rangle+\langle t \rangle$.

As can be easily shown from (\ref{effective_r_m}), (\ref{eq:P_r_mod_main}), and (\ref{estimate_T_r_main}), the resulting efficiency is defined as 

\begin{equation}\label{eq:eta_3}
    \begin{aligned}
        \eta_3 \ge 1 - \frac{{\cal T}_{on}  \langle T^3 \rangle-\langle T^2 \rangle^2}{{\cal T}_{on}  \left[\langle T^3 \rangle - 2 \langle T \rangle \langle T^2 \rangle\right] - 2 {\cal T}_{on} r_3 \left[\langle T \rangle \langle T^3 \rangle-\langle T^2 \rangle^2\right]}.
    \end{aligned}
\end{equation}


%
%

Rate $r_{4}$ minimizes the right side of the inequality (\ref{estimate_T_r_classic}) on the interval (\ref{T_r_range_standart}), thus providing the greatest guaranteed efficiency at the given values of the first three points of the initial completion time

\begin{eqnarray}\label{eq:r^c_opt_main}
    r_4 =\frac{1}{\langle T \rangle}- \frac{\sqrt{\langle T^3 \rangle(\langle T^3 \rangle - 3 \langle T^2 \rangle  \langle T \rangle + 6    \langle T \rangle^2 {\cal T}_{on}})}{\langle T \rangle \langle T^3 \rangle}.
\end{eqnarray}

Efficiency can be estimated from below as

\begin{eqnarray}\label{eq:eta_4}
    \eta_4 \ge 1 - \frac{3 \langle T^2 \rangle - 2 \langle T^3 \rangle r_4}{6 \langle T \rangle \left(\langle t \rangle+\langle T \rangle\right)}.
\end{eqnarray}

The optimal rate is given by

\begin{eqnarray}
    \label{eq:r^ms_opt_main}
    r_5 = \frac{-2 {\cal T}_{on} \langle T^2 \rangle+\sqrt{2} \sqrt{\left(2 {\cal T}_{on}  \langle T \rangle \langle T^2 \rangle+{\cal T}_{on}  \langle T^3 \rangle-\langle T^2 \rangle^2\right)\langle T^2\rangle^3 \left[\langle T \rangle\langle T^3 \rangle\right]^{-1}}}{{\cal T}_{on}\langle T^3 \rangle - \langle T^2 \rangle^2}.
\end{eqnarray}

Efficiency can be estimated from below as

\begin{eqnarray}\label{eq:eta_5}
    \eta_5 \ge 1 - \frac{\langle T^3 \rangle \left(\langle t \rangle+\langle T \rangle\right)-\langle T^2 \rangle^2}{\left(\langle t \rangle+\langle T \rangle\right) \left(\langle T^3 \rangle - 2 \langle T \rangle \left(r_5 \langle T^3 \rangle+\langle T^2 \rangle\right)\right)}.
\end{eqnarray}

\subsection{Optimization of the success probability}
 It is easy to show by examining the right side of the inequality (\ref{p_r_estimate}) at the extremum, the point belonging to the specified interval

\begin{equation}\small
\label{p_r_range}
    r_{0} = \frac{-\langle T_s \rangle \langle T \rangle \langle T^2 \rangle + \sqrt{\langle T \rangle^3 \langle T^2 \rangle \langle T_s \rangle (\langle T \rangle \langle T_s \rangle+\sigma^2)}}{\sigma^2 \langle T^2 \rangle \langle T_s \rangle},
\end{equation} 

where $\sigma^2=\langle T^2\rangle-\langle T\rangle^2$,
provides the maximum guaranteed gain for the given values of $\langle T_s\rangle$, $\langle T\rangle$, and $\langle T^2\rangle$. 
The resulting efficiency is estimated from below as 
\begin{equation}
\begin{aligned}
\label{chi_2}
    \chi_2>& \frac{p}{1 - p} \frac{\langle T \rangle}{\sigma^4} \cdot \biggr[ \left(\left(\langle T \rangle^2+\langle T^2 \rangle\right) \langle T_s \rangle - \langle T \rangle\sigma^2\right)\biggr.\biggr. -2 \sqrt{\langle T \rangle \langle T^2 \rangle \langle T_s \rangle \left(\langle T \rangle \langle T_s \rangle+\sigma^2\right)} \Biggr].
\end{aligned}
\end{equation}

\end{widetext}










\bibliography{Choosing_restart_strategy}

\end{document}